# Infrared echoes near the supernova remnant Cassiopeia A

Oliver Krause[1], George H. Rieke[1], Stephan M. Birkmann[2], Emeric Le Floc'h[1], Karl D. Gordon[1], Eiichi Egami[1], John Bieging[1], John P. Hughes[3], Erick T. Young[1], Joannah L. Hinz[1], Sascha P. Quanz[2] & Dean C. Hines[4]

[1]Steward Observatory, University of Arizona, 933 N. Cherry Avenue, Tucson, AZ 85721, USA

[2]Max-Planck-Institut für Astronomie, Königstuhl 17, 69117 Heidelberg, Germany

[3]Department of Physics & Astronomy, Rutgers University, 136 Frelinghuysen Road, Piscataway, NJ 08854, USA

[4]Space Science Institute, 4750 Walnut Street, Boulder, CO 80301, USA

**Two images of Cassiopeia A (Cas A) obtained at 24 μm with the *Spitzer* Space Telescope over a one year time interval show moving structures outside the shell of the supernova remnant to a distance of more than 20 arcmin. Individual features exhibit apparent motions of 10 to 20 arcsec per year, independently confirmed by near-infrared observations. The observed tangential velocities are at roughly the speed of light. It is likely that the moving structures are infrared echoes, in which interstellar dust is heated by the explosion and by flares from the compact object near the center of the remnant.**

Cas A is the youngest supernova remnant (SNR) known in our Galaxy, thought to arise from the core-collapse of a massive (~ 20 $M_\odot$) Wolf-Rayet star (*1*) about 325 years ago (*2*). Its youth and proximity to Earth ($3.4^{+0.3}_{-0.1}$ kpc) (*3*) make Cas A an unique "laboratory" for supernova astrophysics. We imaged Cas A at 24, 70, and 160 µm on 30 November 2003 (*4,5*) using the Multiband Imaging Photometer for *Spitzer* (MIPS) (*6*) in scan map mode. The observations yielded a rectangular image from which a section about 40 arcmin long by 10 arcmin wide is displayed in Fig. 1A. At the distance of Cas A, 1.0 arcmin corresponds to 1.0 pc or about 3 light years. The 24 µm image revealed fine filaments and knots that stand out clearly against the more diffuse galactic cirrus emission around the remnant. Two bright lobes of a bipolar structure are axisymmetrically located 14 arcmin NE and 11 arcmin SW from the center of the remnant (Fig. 1A).

We obtained near infrared ($\lambda$ = 2.2 µm) images of the northern lobe using the Multiple Mirror Telescope (MMT) on Mt. Hopkins (Arizona) on 30 May 2004 and the Calar Alto 3.5m telescope (Spain) on 11 October 2004 and 8 January 2005. All three images show faint filamentary nebulosities within the region of the brightest 24 µm emission. These nebulosities undergo remarkable changes in morphology over the time interval of only a few months. One such filament is shown in Fig. 2, the emission region of which is moving at an angular tangential velocity of 18 ± 2 arcsec/yr, close to the speed of light at the distance of the SNR.

We repeated our initial *Spitzer* scan map at 24 µm a year later, on 2 December 2004. The difference image in Fig. 1B shows dramatic changes out to the edge of our images about 25 arcmin (80 light years) from the SNR. The two images within the northern lobe region in Fig. 3 are typical of the huge variations. The morphological variations make it difficult to follow distinct filaments over the one year time interval; however, there is a persistent ~ 7 arcmin long filament (see Fig. 1B) connecting the southern lobe of the bipolar structure with the remnant whose cusp exhibits a tangential velocity of $v \sim 0.7 \pm 0.1$ c. There is a marked tendency for the motions in the bipolar structure to be outward from the center of the SNR. Our K-band images also show a number of features outside the bipolar structure with apparent motions in other directions.

High-velocity ejecta have previously been identified in optical emission lines (*7*) out to 4.5 arcmin distance and with expansion speeds up to nearly 15,000 km/s. If the infrared features were produced by ejecta from the supernova explosion 325 years ago, their location would require average space velocities of at least a quarter of the speed of light, or 75,000 km/s, in addition to the observed apparent velocities close to the speed of light. Such velocities have never been observed from any individual ejecta in supernova remnants, nor have they been spectroscopically inferred from supernova outbursts. Apparent motions close to the speed of light have however been observed in scattered light and infrared echoes originating from supernova explosions (*8,9*) and other brilliant light sources (*10,11*).

We believe that infrared echoes, i.e. the heating of interstellar dust by the outward moving photon shell of a bright flash and the echoed re-radiation of the resulting infrared emission (*12*), provide the most convincing interpretation for the moving infrared features near Cas A. This hypothesis is supported by images of the northern lobe at 3.6, 4.5, 5.8 and 8 µm obtained with *Spitzer* on 17 December 2004 and at 2.2 µm from Calar Alto on 8 January 2005, to accompany the December 2 ones at 24 µm. The morphologies in all these bands are similar, allowing us to identify individual features reliably. The average spectral energy distribution (SED) shown in Fig. 4 is consistent with emission dominated by a smooth, but very red, continuum (but with some spectral structure across the 3.6 – 8 µm region) and suggestive of thermal emission from warm dust. The reddening toward Cas A ($A_K \sim 1$) is insufficient to change the overall shape of the SED substantially. A direct scattered light echo is unlikely since it would be blue in color (*8*), in contrast to the observed red continuum.

A thermal dust continuum is also consistent with the lack of line emission ($3\sigma$ upper limit of 60 µJy; 1 Jy = $10^{-26}$ W m$^{-2}$ Hz$^{-1}$) in K-Band spectroscopy (spectral resolution $\lambda/\Delta\lambda \sim 500$) of two bright emission peaks obtained on 30 July 2004. Line emission is also not detected in narrow band images of the northern lobe region in filters for $H_2$ (2.12 µm) and Br $\gamma$ emission obtained on 8/11 October 2004. We can identify the lobe to the SW in our 70 µm image at a level consistent with thermal emission. However, confusion with bright galactic cirrus emission and the lower spatial resolution at this wavelength complicate an unambiguous determination of the 70 µm flux density for the moving features. A bright filament wisp is faintly detected in a deep R band image obtained on 27 December 2003; because of the uncertainty in the extinction, we have not tried to include this point in the SED.

Emission by an infrared echo is also consistent with the non-detection of the filaments in the radio continuum and at X-ray energies. No trace of the filaments was detected at cm-wavelengths toward three fields shown in Fig. 1, using the Very Large Array (VLA) in BnA configuration at C-band (4.9 GHz) and X-band (8.4 GHz) on 6/7 February 2005. For these fields, from north to south, the rms noise levels at C band were 40, 66, and 80 µJy (beam size $0.9 \times 0.5$ arcsec$^2$ FWHM); and at X band were 56, 60, and 81 µJy (beam size $1.6 \times 0.9$ arcsec$^2$). This is far below (less than one thousandth) a traditional synchrotron power law spectrum (spectral index -0.7) normalized at 24 µm and argues against a relativistic jet being responsible for the apparent superluminal motions (*13*). The Chandra High Resolution Camera (HRC) imaged Cas A on 19 December 1999. The field of view covered the two inner circled regions in Fig. 1 about 12 arcmin to the north and south of the SNR. No significant X-ray emission was detected within either of the circled regions to an unabsorbed flux limit of $1 \times 10^{-13}$ erg cm$^{-2}$ s$^{-1}$ (0.2--6 keV band). Given the variability of these features, it would be desirable to confirm this result with more contemporaneous X-ray and infrared measurements.

If the heating flash for the infrared echoes was emitted by the supernova explosion, then the structures we see now must lie 50 - 60pc behind the remnant to account for the delay of 325 years in light travel time. The angle of incidence of the flash relative to the plane of the sky would be about 75º, and as a result there would only be a mild tendency for an infrared echo to prefer outward motion. While some of the infrared features with non-systematic motions may arise from dust heated by the initial explosion, the strong observed outward tendency of the infrared echoes in the bipolar structure suggests the presence of an object within the remnant that has recently had short outbursts. The sharp appearance of the infrared features, which are partly unresolved in our K-band images at 0.7 arcsec resolution, indicates that the time scale of such an outburst and the subsequent dust cooling can be no more than a few weeks (The expected cooling time scales for typical interstellar dust particles are significantly shorter than this limit (*14*)). The geometry of the echoes is consistent with heating by short flares of beamed light emission roughly perpendicular to the line of sight from the central compact X-ray source (*15*) whose precise nature is unknown. Assuming that the two lobes NE and SW of the remnant correspond to infrared echoes of one such beamed flare, we can determine its date and space direction from the position of the sharply defined outer lobe boundaries relative to the central object. Equating the light travel time to the NE lobe (14.0 ± 0.5 arcmin) and the SW one (11.3 ± 0.5 arcmin) yields a flare date of AD 1952.9 ± 2.5, at position angle 26 ± 4º on the sky and at an angle of 82 ± 3º relative to the line of sight. The flare may have escaped direct detection because it was beamed nearly perpendicular to the line of sight.

Strongly magnetized neutron stars (magnetars) localized in young supernova remnants (*16*) are known to emit short, non-periodic, and luminous flares primarily detected at high photon energies (*17*), in which case they are described as soft gamma ray repeaters (SGRs). Giant flares from SGRs are also accompanied by outbursts of relativistic particles (*18,19*). The position angle of the AD 1952 flare coincides with the location (position angle ~ 26º) of the radio structure H (*20*), which is one of the outermost bright features of Cas A. Radio observations in 1974 revealed a compact radio knot within this structure (*21*), which was not visible in 1969. In case the enhanced synchrotron emission of this feature is powered by relativistic particles from the ~ AD 1952 flare, their average space velocity would have to be $v$ ~ 0.6 c to have reached the position of the radio knot between 1969 and 1974. This is consistent with the inferred velocity of ablated baryons in a recent SGR flare (*19*).

The X-ray point source near the center of the remnant is believed to be a neutron star. Although no soft gamma ray bursts have ever been detected from it, its spectral properties are consistent with a momentarily quiescent SGR (*22*). Alternatively, it has been suggested that the central object is in the process of evolving into a SGR (*23*). While the status of this object remains uncertain, we take the properties of SGRs as a guideline in a rough calculation of luminosity. SGR giant flares can have isotropic energy equivalents of $10^{44}$ to $5 \times 10^{46}$ ergs (*17, 24*); however, there are indications that their emission might be anisotropic (*25*). The bolometric IR luminosity of the features in the northern lobe region is ~ 0.02 $L_\odot$ arcsec$^{-2}$, or about 0.6 $L_\odot$ within our 6 arcsec (FWHM) beam at 24 µm. We hypothesize that the infrared features represent the effect of a light pulse passing through the interstellar gas, which instantaneously heats dust it encounters along an expanding sphere centered on the neutron star. Assuming the northern lobe to represent a tangent position to this sphere, the time for the heated region to transit the 6 arcsec beam is about 4 months. A flare of isotropic energy equivalent $2 \times 10^{46}$ ergs absorbed entirely over this time period would yield a net luminosity of ~ 60 $L_\odot$ within the 24 µm beam, about 100 times more than observed. The luminosities of SGRs therefore appear consistent with the heating requirements for the infrared features, even accounting for a possibly low filling factor of the absorbing material.

26. We appreciate allocations of Director's Discretionary Time on *Spitzer*, the VLA and the Calar Alto observatory. We thank Mark Claussen for advice on the VLA observing strategy. We thank Don McCarthy for use of the PISCES camera. This research was supported by NASA through contract 960785 issued by JPL/Caltech. The VLA is a facility of the National Radio Astronomy Observatory, which is operated by Associated Universities, Inc. under contract with the National Science Foundation.

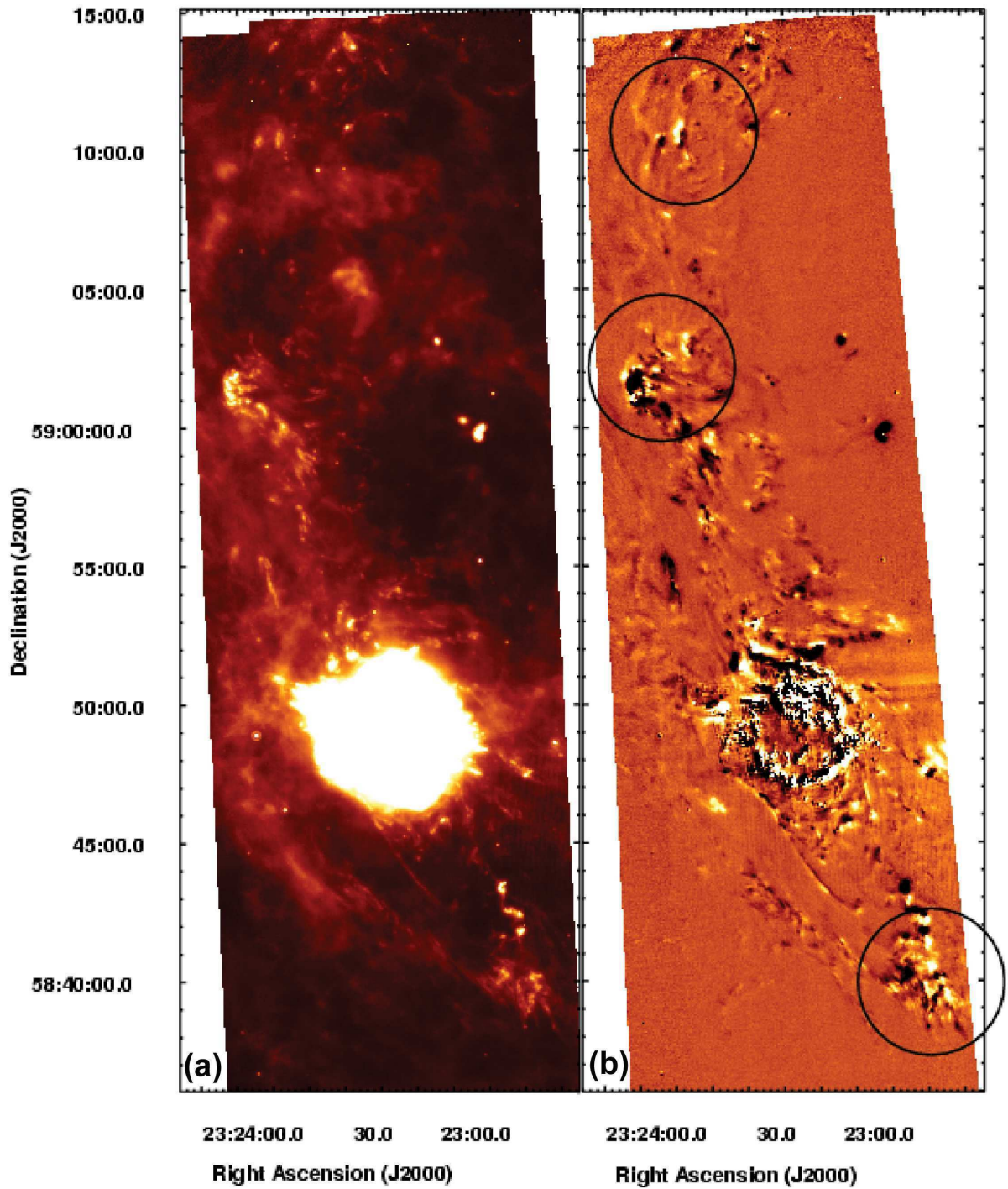

Fig. 1.— (a): MIPS scan map image of Cas A obtained on 30 November 2003. Colors represent 24 μm surface brightness (range 18-35 MJy/sr); (b): Difference image of Cas A at 24 μm. Black features correspond to the MIPS scan map on 30 November 2003 and white to observations on 2 December 2004. The location and size of the VLA fields are indicated by circles. Note that the image components due to the stationary IR cirrus cancel out very well in the difference image (e.g., to the NW and SE of the SNR).

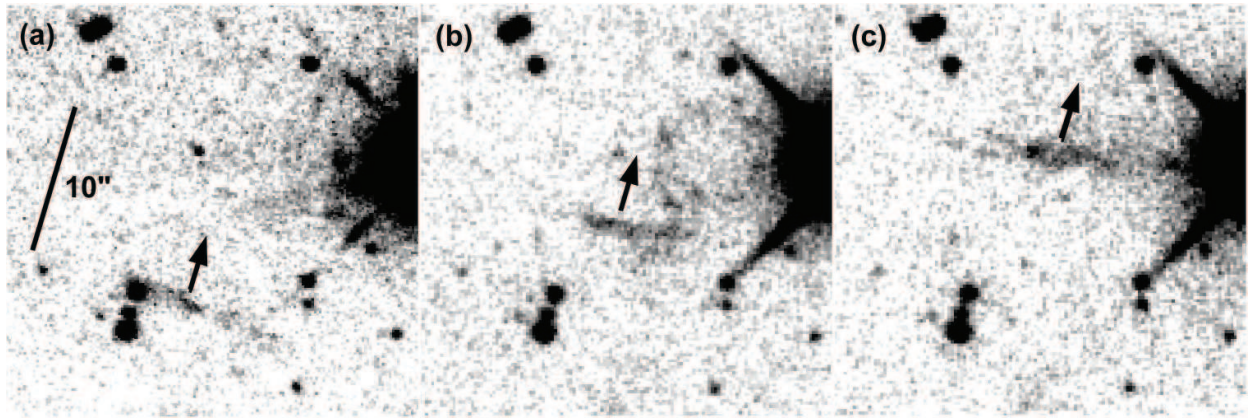

Fig. 2.— $K_s$-band images of a moving filament in the northern lobe region obtained in May 2004 from MMT (A) and in October 2004 (B) and January 2005 (C) from Calar Alto (from left to right). The field size is 30×30 arcsec$^2$. The 10 arcsec scale in the left panel corresponds roughly to the light travel distance during the ∼ seven month interval shown here. The seeing for all three images was about 0.7 arcsec.

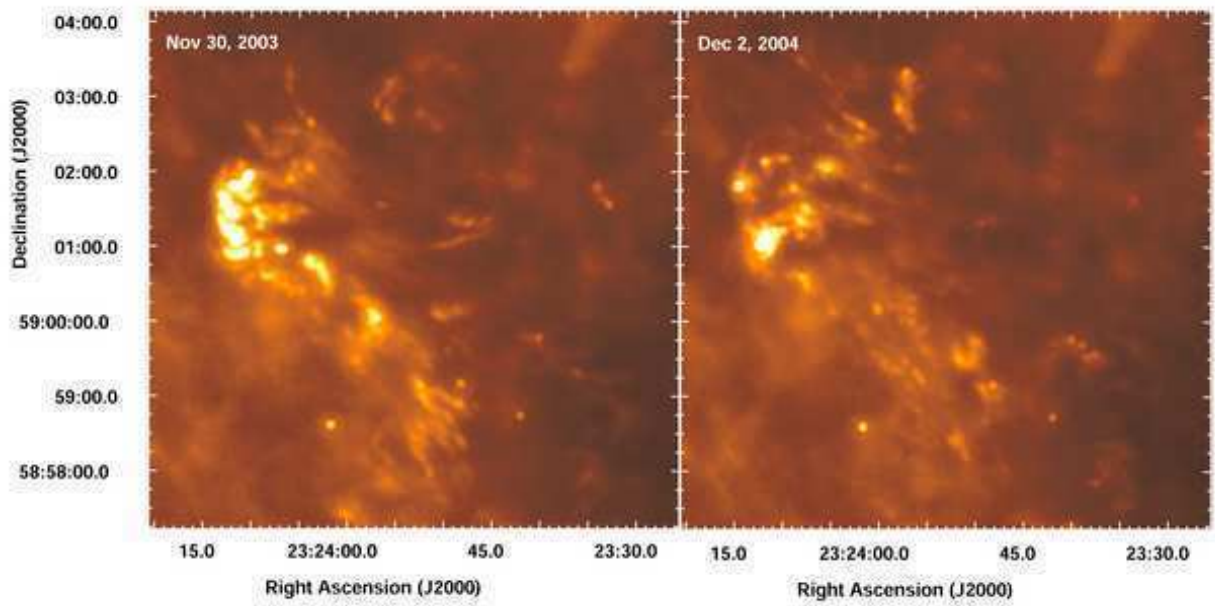

Fig. 3.— 24 $\mu$m MIPS images of the northern lobe region. Large proper motions and morphological variations occurred in the filamentary nebulosity over the 1 year interval.

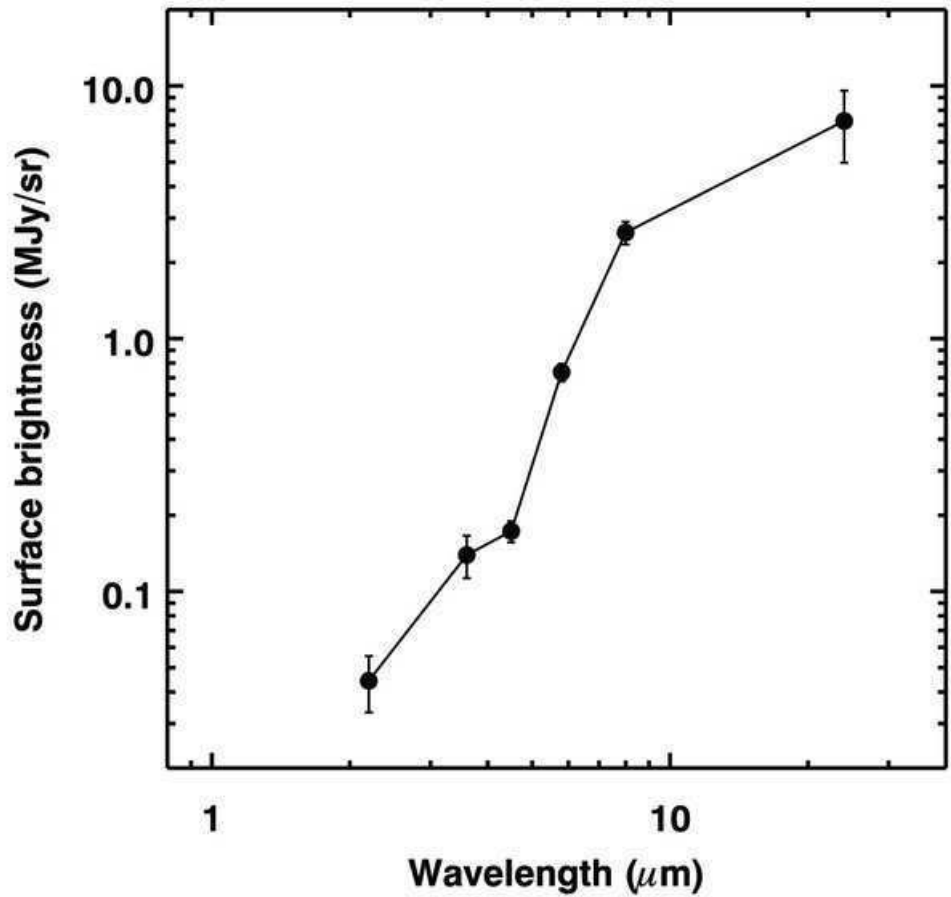

Fig. 4.— Average spectral energy distribution of the infrared features in the northern lobe region. Images at 3.6, 4.5, 5.8 and 8 $\mu$m from *Spitzer* and at 2.2 $\mu$m from Calar Alto were convolved to match the MIPS 24 $\mu$m beam. The surface brightness was then measured point-by-point at 59 positions in Nyquist sampled images after removal of field stars and diffuse cirrus emission. The SEDs for each of the statistically independent positions were normalized to their average. The quoted 1 - $\sigma$ uncertainties include the standard deviation of the individual normalized data points in each band and their systematic calibration errors